\documentstyle[12pt]{article}

\newcounter{tabl}
\newcommand{\be}{\begin{equation}}
\newcommand{\ee}{\end{equation}}
\newcommand{\beq}{\begin{eqnarray}}
\newcommand{\eeq}{\end{eqnarray}}
\newcommand{\bea}[2]{\be\label{#2}\begin{array}{#1}}
\newcommand{\eea}{\end{array}\ee}

\def\Nb{{\rm \bf N}}
\def\Rb{{\rm \bf R}}
\def\Cb{{\rm \bf C}}

\def\det{\,{\rm det}\, }

\def\tr{\,{\rm tr}\, }

\def\sign{{\rm sign}}

\def\({\left(}
\def\){\right)}
\def\[{\left[}
\def\]{\right]}
\def\p{\partial}

\def\11{1\!\! 1}

\def\hf{{1\over 2}}

\def\eps{\varepsilon}

   \def\CC {{\cal C}}

   \def\CG {{\cal G}}
   \def\CH {{\cal H}}
   \def\CI {{\cal I}}
   
   \def\CK {{\cal K}}
   \def\CL {{\cal L}}
   \def\CM {{\cal M}}

   \def\CP {{\cal P}}

   \def\CS {{\cal S}}

   \def\CV {{\cal V}}
   \def\CW {{\cal W}}

\newcommand{\tE}{\lefteqn{\smash{\mathop{\vphantom{<}}\limits^{\;\sim}}}E}
\newcommand{\tP}{\lefteqn{\smash{\mathop{\vphantom{<}}\limits^{\;\sim}}}P}
\newcommand{\tQ}{\lefteqn{\smash{\mathop{\vphantom{<}}\limits^{\;\sim}}}Q}
\newcommand{\Et}{\lefteqn{\smash{\mathop{\vphantom{\Bigl(}}\limits_{\sim}
\atop \ }}E}
\newcommand{\Pt}{\lefteqn{\smash{\mathop{\vphantom{\Bigl(}}\limits_{\sim}
\atop \ }}P}
\newcommand{\Qt}{\lefteqn{\smash{\mathop{\vphantom{\Bigl(}}\limits_{\sim}
\atop \ }}Q}

\newcommand{\tN}{\lefteqn{\smash{\mathop{\vphantom{\Bigl(}}\limits_{\sim}
\atop \ }}N}

\newcommand{\tNn}{\lefteqn{\smash{\mathop{\vphantom{\Bigl(}}\limits_{\,\sim}
\atop \ }}{\cal N}}

\newcommand{\gl}{{\rm g}}
\newcommand{\SA}{{\cal A}}

\newcommand{\tSA}{{\widetilde \SA}}

\newcommand{\tPb}{{\tP_{\smash{(\im)}}}}

\newcommand{\nd}{{\cal N}_D}

\newcommand{\IQ}{{I_{\smash{(Q)}}}}

\newcommand{\im}{\beta}

\newcommand{\SLR}{{\rm SL(2,\Rb)}}
\newcommand{\SLC}{{\rm SL(2,\Cb)}}
\newcommand{\SUU}{{\rm SU(2)}}
\newcommand{\sgchi}{${\rm SU}_{\chi}(2)$ }

\newcommand{\Ppr}[2]{\CI^{(#1)}{(#2)}}

\newcommand{\slqg}{SLQG}

\newcommand{\Ref}[1]{(\ref{#1})}

\def\plabel#1{\label{#1}}
 
\begin{document} 
%
%

\title{ \Large \bf
Timelike surfaces in Lorentz covariant loop gravity \\
and spin foam models}

\author{Sergei Alexandrov\thanks{email: S.Alexandrov@phys.uu.nl}
\and Zolt\'an K\'ad\'ar\thanks{email: Z.Kadar@phys.uu.nl}}

\date{}

\maketitle

\vspace{-0.8cm}

\begin{center}
\it  Institute for Theoretical Physics \& Spinoza Institute, \\
Utrecht University, Postbus 80.195, 3508 TD Utrecht, The Netherlands
\end{center}

\vspace{0.1cm}

\begin{abstract}
We construct a canonical formulation of general relativity for the case of a timelike
foliation of spacetime. The formulation possesses explicit covariance with respect to Lorentz 
transformations in the tangent space. Applying the loop approach to quantize the theory
we derive the spectrum of the area operator of a two-dimensional surface. 
Its different branches are naturally associated to spacelike and timelike surfaces.
The results are compared with the predictions of Lorentzian spin foam models.
A restriction of the representations labeling spin networks leads to 
perfect agreement between the states as well as the area spectra in the two approaches.   
\end{abstract}

\section{Introduction}

There are many approaches to quantization of gravity
(for a recent review, see \cite{Carlip}). One of the promising
approaches is the idea of loop quantization \cite{loops1,loops2}.
It suggests that excitations of quantum space are concentrated on one-dimensional
structures like loops or graphs which establish relations between different 
points called vertices. Developing this picture of quantum space in time,
one obtains a representation of quantum spacetime as a complex of branched
surfaces. 

The latter picture appears in another approach to quantum gravity known as
spin foam models \cite{Oriti,perez}. These models realize the idea that quantum gravity
can be obtained as a sum over histories of quantum spacetime which is
a generalization of the usual path integral quantization. Thus, the two
different ideas for quantizing gravity lead to the same qualitative picture.
Do they agree quantitatively?
It turns out that the answer to this question is in the negative. 
In fact, the origin of the disagreement is easily traced back to 
the starting points of the two approaches. 

The standard loop quantization is based on
the so-called Ashtekar--Barbero formalism and leads to the theory which we call
SU(2) Loop Quantum Gravity (\slqg) (for review, see \cite{Rov,Rov-dif}).
It starts with the first order formulation of general relativity in $3+1$
dimensions with the Lorentz gauge group in the tangent space.
Then, as a result of a partial gauge fixing, the canonical formulation 
possesses, besides the usual diffeomorphism invariance, only a local
SU(2) symmetry. Choosing the SU(2) connection as one of the canonical variables,
one can construct loop variables from Wilson lines defined by this connection.
These variables have a simple loop algebra and the construction of
the (kinematical) Hilbert space is then straightforward. In particular,
an orthonormal basis is realized by the so-called spin network states
constructed from SU(2) Wilson lines in irreducible SU(2) representations.

On the other hand, spin foam models of Lorentzian general relativity 
do not break the covariance in the tangent space and 
essentially use the representation theory of the Lorentz group
\cite{BC,sfLor1,sfLor2}.
Therefore, their predictions involve Lorentz, rather than SU(2), structures.
For example, as it was shown recently in detail in \cite{Maran3}, 
if one takes a slice of a spin foam by a 3-dimensional hypersurface,
the spin foam induces a spin network state on the slice. But its elements
(edges and vertices) are labeled by irreducible representations of \SLC.
Thus, \slqg\ and the spin foam models clearly differ at the quantitative level.

Although there were some attempts to find an agreement \cite{Maran1,Maran2},
they cannot reach a full success since these two approaches are
based on different structures. Therefore, either one of them or both should be
modified if we expect that the agreement on the qualitative level is not accidental.
Of course, the modern spin foam models do not have a rigorous ground and represent 
in some sense just a reasonable discretization of the path integral.
So it would not be a surprise that some modification of them will be required.  
However, it is very unlikely that such a modification will reduce the gauge symmetry
and replace everywhere \SLC\ by \SUU.
Instead, as we will argue below, it is \slqg\ that requires to be seriously
reconsidered.

A first sign of this is that \slqg\ suffers from several problems. 
Besides the problem of the absence of Lorentz invariance, which is the core
of the disagreement with the spin foam models, the most evident one is the so-called
Immirzi parameter problem \cite{imir}.
It refers to the main result of \slqg, which is
the spectrum of the area operator of two-dimensional spacelike surfaces \cite{area1,ALarea}.
It was shown that the spectrum is given by the sum of square roots of 
the SU(2) Casimir operator over punctures of the measured surface
by a spin network.
The problem is that this result depends also on a non-physical parameter,
which is called Immirzi parameter. In the classical theory this parameter can be freely
introduced without changing the equations of motion. But at the quantum level
in the SU(2) loop approach, it affects all results what indicates that some
anomaly is present.

It was thought that the anomaly in question is a physical one so that
the Immirzi parameter becomes a new fundamental constant.
However, in a series of works \cite{SA,AV,SAcon,SAhil,AlLiv} it was shown
that this is not the case because it is actually a consequence of a diffeomorphism anomaly, 
whereas there is a quantization which preserves all classical 
symmetries and leads to results independent of the Immirzi parameter. 

This quantization is based on a Lorentz covariant canonical formulation of 
general relativity following from the first order formalism if one does not fix any gauge 
\cite{SA}. In fact, the loop quantization of this formulation is not unique 
and depends on the choice of variables one uses to define Wilson line operator.
In particular, it was shown that there exists
a two-parameter family of Lorentz connections such that
the area operator is diagonal on the Wilson lines defined by them \cite{SAcon}. 
All of these connections lead to
different area spectra so that the choice of the connection to be used for
quantization represents a real quantization ambiguity of this approach.

This ambiguity was fixed by requiring the correct transformation properties
under time diffeomorphisms. It was shown that there is a unique connection from
the family satisfying this condition. The area spectrum corresponding to it is
expressed through the difference of the \SLC\ and \SUU\ Casimir operators
and it does not depend on the Immirzi parameter \cite{AV,SAcon}.

On the other hand, it turned out to be possible to derive \slqg\ from the covariant
quantization. Namely, it corresponds to a choice of connection
from the two-parameter family, which is different from the one
mentioned in the previous paragraph that ensures
the correct transformation properties.
But since the connection does not transform correctly
under all symmetries of the classical theory, it was argued that
the quantization breaks the diffeomorphism invariance \cite{SAcon,AlLiv}.
In particular, this can explain the appearance of the Immirzi parameter in
the physical quantities.

The described results imply that the unique way to proceed in the loop approach is
to use the connection ensuring all symmetries to be preserved after quantization.
We call the resulting theory Covariant Loop Quantum Gravity (CLQG).
We emphasize that it is not just an alternative quantization but
it predicts that \slqg\ is not correct.

Let us emphasize however that CQLG is far from being completed.
In the covariant approach the Wilson line operators belong to the 
Lorentz group and the fact that it is non-compact essentially complicates the quantization.
But even a more serious obstacle is the presence of the second class constraints.
Due to these reasons, even the construction of the kinematical Hilbert space of CLQG
was not completed although there were several proposals in this direction
\cite{SAhil,AlLiv}. 

Nevertheless, the presence of the full Lorentz symmetry suggests the possibility 
to establish a connection with the Lorentzian spin foam models. Some similarities 
have been already observed in the proposals for the Hilbert space \cite{SAhil,AlLiv}.
Moreover, in \cite{AlLiv} it was shown how one can reproduce the states induced 
by spin foams from the states of CLQG. Thus, the connection of CLQG and spin 
foam models is becoming tighter.

In this paper we push forward the relation between the two approaches. 
It is known that the faces of a Lorentzian spin foam are labeled with irreducible 
representations of the Lorentz group either from the continuous series $(0,\rho)$
or from the discrete ones $(n,0)$. The former give rise to spacelike surfaces,
whereas the latter define timelike surfaces because the corresponding area
is either real or imaginary \cite{BC}. However, in the usual canonical approach 
one cannot measure the area of a timelike surface since it is not embedded into one 
slice of the pre-defined foliation. Thus, the loop quantization is able to capture 
only the first class of surfaces.
In particular, the area operator of \cite{AV,SAcon} is always real confirming
this expectation. 

Here we show how timelike surfaces can still be incorporated into the 
framework of CLQG and that the resulting spectrum of the area operator agrees
with predictions of the spin foam approach.
The idea is to use a foliation of spacetime 
with timelike, rather than spacelike, slices to define the Hamiltonian formulation.
Of course, it is not evident at all that a quantum theory based on a formulation, 
where the role of time is played by a spacelike coordinate, can be meaningful. 
However, there is a hope to capture at least some local properties
of the real quantum theory. Moreover, in quantum gravity the causal structure
is expected to be fluctuating. Therefore, one cannot guarantee for 
any pre-defined foliation to have fixed causal properties. 

Thus, without caring much about the meaning of the resulting theory,
we perform a loop quantization of general relativity defined on a timelike foliation. 
It turns out that, working with the Lorentz covariant formulation of \cite{SA}, 
almost nothing should be changed if we want to describe a timelike instead of a spacelike
foliation. This allows to avoid any calculations because all of them can be borrowed from 
the previous
works. As a result, one arrives at the area spectrum in a few steps and finds
that it has the same structure as the previous one \cite{AV,SAcon} with the only difference 
that the Casimir operator of \SUU\ is replaced by the one of \SLR.

We should note that the idea to use a temporal foliation already appeared in
the literature \cite{Maran2}. However, our analysis is much simpler and leads directly
to results which can be identified with those of the spin foam models. In particular, we
show that a certain restriction of representations labeling the so-called
projected spin networks, which provide an orthonormal set of states in the (extended) 
kinematical Hilbert space, gives the states induced by spin foams on temporal slices.
 
The structure of the paper is the following. In the next section we 
review the Lorentz covariant canonical formulation and also present some new results
on the Dirac algebra of canonical variables.
An important new observation is that the dependence of the Immirzi parameter completely
disappears from the Dirac brackets of the elementary variables, a triad multiplet and 
the spacetime connection, which are used to construct quantum operators. 
In section 3 we show how the case of a timelike foliation
can be incorporated and what is the corresponding area operator. Then in section 4
we recall the notion of projected spin networks and 
establish a relation with the spin foam models.
Finally, section 5 is devoted to the discussion.

We use the following notations.
The indices $\mu,\nu,\dots$ from the middle of the Greek alphabet
are used to label spacetime coordinates and $\alpha,\beta,\dots$
from the beginning of the alphabet denote Lorentz
indices in the tangent space. The $3+1$ decomposition is done according to 
the following notations: $\mu=(0,i)$ and
$\alpha=(0,a)$, so latin indices
$i,j,\dots$ from the middle of the alphabet label the space coordinates and $a,b,\dots$ 
from the beginning are the $so(3)$ indices. The capitalized latin indices $X,Y,,\dots$
take 6 values and label the components of the adjoint representation of $sl(2,\Cb)$
(which correspond to the antisymmetrized pairs $\alpha\beta$).
The signature of the metric is assumed to be $(-,+,+,+)$ and
the Levi-Civita symbol is normalized as $\eps_{0123}=1$.

\section{Lorentz covariant canonical formulation}

\subsection{Decomposed action and canonical analysis}

The Lorentz covariant canonical formulation comes from the $3+1$ decomposition of
the generalized Hilbert--Palatini action
\be
S_{(\beta)}=\frac12 \int \eps_{\alpha\beta\gamma\delta}
e^\alpha \wedge e^\beta \wedge (\Omega^{\gamma\delta}+
\frac{1}{\beta}\star\Omega^{\gamma\delta}),
\plabel{palat}
\ee
where $e^\alpha$ is the tetrad field,
$\Omega^{\alpha\beta}$ is the curvature of
the spin-connection $\omega^{\alpha\beta}$ and
the star is the Hodge operator defined as
$ \star \Omega^{\alpha\beta}=\frac 12 {\eps^{\alpha\beta}}_{\gamma\delta}
\Omega^{\gamma\delta} $.
The parameter $\beta$ coincides with the Immirzi parameter
and is not physical since the
additional term in the action is purely topological and does not change
the equations of motion.

The $3+1$ decomposition of the tetrad reads as follows
\be
e^0=Ndt+\chi_a E_i^a dx^i ,\qquad e^a=E^a_idx^i+E^a_iN^idt.
\plabel{tetrad}
\ee
The distinction from the decomposition used in \slqg\ is the presence
of the field $\chi$. It is absent in the Ashtekar--Barbero formalism
since the condition $\chi=0$ called ``time gauge" is imposed from the very beginning.
This corresponds to fixing of the gauge freedom related to Lorentz boosts
in the tangent space. As usual, it is convenient to introduce 
the inverse triad $E^i_a$ and densitized fields
\be
\tE^i_a =h^{1/2}E^i_a,  \qquad
\tN=h^{-1/2} N, \qquad \sqrt{h}=\det E^a_i.
\plabel{densit}
\ee
To write the decomposed action we also need to redefine the lapse and shift
variables
\be
N^i=\nd^i+\tE^i_a\chi^a\tNn, \quad
\tN=\tNn+\Et_i^a\chi_a\nd^i
\ee
and to introduce fields in the adjoint representation of the Lorentz group
\beq
& A_i^{ X}=(\omega_i^{0a},\frac12 {\eps^a}_{bc}\omega_i^{bc}),
     &
\nonumber\\
&  \tP_X^{ i}=(\tE^i_a,{\eps_a}^{bc}\tE^i_b\chi_c),
\qquad 
\tQ_X^{ i}=(-{\eps_a}^{bc}\tE^i_b\chi_c,\tE^i_a),
     &
\plabel{multHP}\\
&  \tPb_X^i=\tP_X^i-\frac{1}{\beta}\tQ_X^i.
     &
\nonumber
\eeq
The index $X$ can be thought as an antisymmetric pair $(\alpha\beta)$.
Then the first 3 components correspond to $(0,a)$ and the other 3 are obtained
after contraction of $(ab)$ components with $\hf \eps^{abc}$.
Thus, the first field is just the space components of the spin-connection
$\omega^{\alpha\beta}$. The second field can be obtained from the
bivector $e^{\alpha}\wedge e^{\beta}$, whereas the third one comes from its
Hodge dual. This fact is reflected in the relation
\be
\tP^i_X=\Pi_X^Y\tQ^i_Y,  \qquad
\Pi^{XY} =g^{XZ}\Pi_Z^Y=\left(
\begin{array}{cc}
0&1 \\ 1&0
\end{array}
\right)\delta^{ab}, \qquad
g^{XY}=\left(
\begin{array}{cc}
1&0 \\ 0&-1
\end{array}
\right)\delta^{ab},
\plabel{relPQ}
\ee
where the matrix $\Pi^{XY}$ is a representation of the Hodge operator
and we used the Killing form $g^{XY}$ of the $sl(2,\Cb)$ algebra to raise indices
in the adjoint representation.
Finally, the last definition in \Ref{multHP} suggests to introduce
\be
R^{XY} =g^{XY}-\frac{1}{\beta}\Pi^{XY}=\left(
\begin{array}{cc}
1& -\frac{1}{\beta} \\
 -\frac{1}{\beta} & -1
\end{array}
\right)\delta^{ab} .
\plabel{matR}
\ee
Some properties of the matrices $\Pi$ and $R$ and of the structure constants
$f_{XY}^Z$ of the Lorentz algebra are presented in Appendix A.

In terms of the introduced fields the decomposed action takes the following form
\cite{SA}
\beq
S_{(\beta)} &=&\int dt\, d^3 x (\tPb^i_X\partial_t A^X_i
+A_{0}^X \CG_X+\nd^i H_i+\tNn H),
\plabel{dact}  \\
\CG_X&=&\partial_i \tPb^i_X +f_{XY}^Z A^Y_i \tPb^i_Z,
\plabel{Gauss}\nonumber \\
H_i&=&-\tPb^j_X F_{ij}^X,
\plabel{hdiff}\nonumber \\
H&=&-\frac{1}{2\left(1+\frac{1}{\beta^2}\right)}
\tPb^i_X \tPb^j_Y f^{XY}_Z R^Z_W F_{ij}^W,
\plabel{ham} \nonumber \\
F^X_{ij}&=&\partial_i A_j^X-
\partial_j A_i^X+f_{YZ}^X A^Y_i A^Z_j.
\plabel{FF} \nonumber
\eeq
This action resembles the action of the Ashtekar--Barbero formalism.
There are ten first class constraints $\CG_X,\ H_i$ and $H$ corresponding
to local Lorentz and diffeomorphism symmetries, and one of the canonical variables
is a gauge connection. However, the canonical structure of the action \Ref{dact}
is much more complicated due to the presence of second class constraints.
There are two sets of such constraints:
\beq
\phi^{ij}&=&\Pi^{XY}\tQ^i_X\tQ^j_Y=0,
\plabel{phi} \\
\psi^{ij}&=&2f^{XYZ}\tQ_X^{l}\tQ_Y^{\{ j}\partial_l \tQ_Z^{i\} }
-2(\tQ\tQ)^{ ij }\tQ_Z^{l}A_l^Z+
2(\tQ\tQ)^{l\{i  }\tQ_Z^{j\}}A_l^Z = 0.
\plabel{psi}
\eeq
They require a modification of the symplectic structure to that of the Dirac brackets.
As a result, the canonical variables acquire non-trivial commutation relations.
The details of the canonical analysis can be found in \cite{SA}.

\subsection{Lorentz connections, Dirac algebra and area spectrum}

If one tries to use the canonical formulation described as a starting point
for the loop quantization, one encounters an immediate problem.
It is easy to define Wilson lines of the canonical connection $A^X_i$
and the area operator constructed from $\tP^i_X$. However, due to the modified
commutation relations the action of the latter is not diagonal on such Wilson lines.
This problem was solved in \cite{AV,SAcon} where it was shown that one can
shift the canonical connection in such a way that the Wilson lines of the shifted
connection are eigenstates of the area operator. The additional requirement of the correct
transformation properties under {\it all} classical symmetries led to a unique connection

\be
\SA_i^X=A_i^X + \frac{1}{2\left(1+\frac{1}{\im^2}\right)}
R^{X}_{S}I_{(Q)}^{ST}R_T^Z f^Y_{ZW}\Pt_i^W \CG_Y.
\plabel{spcon}
\ee      

At this point we have to introduce new fields appearing in \Ref{spcon},
the inverse triad multiplets, $\Pt_i^X$ and
$\Qt_i^X$, satisfying
\be
 \tQ^i_X\Qt_j^X=\delta^i_j,  \qquad
\tP^i_X\Pt_j^X=\delta^i_j, \qquad \tQ^i_X\Pt_j^X=\tP^i_X\Qt_j^X=0, 
\ee
and the projectors
\be
I_{(P)X}^Y = \tP^{ i}_X\Pt_i^{Y},
\qquad
I_{(Q)X}^Y = \tQ^{ i}_X\Qt_i^{Y}.
\plabel{proj}
\ee
These projectors will play an important role in the following, therefore
we explain their main properties. The explicit expressions of all these fields in
terms of the triad and $\chi$ can be found in \cite{SA}.
In particular, the projectors \Ref{proj}
are constructed from the field $\chi$ only.
The name ``projector" for the quantities \Ref{proj} is justified by the following
relations
\beq
& I_{(P)X}^Y +I_{(Q)X}^Y=\delta_X^Y, &
\plabel{projdel} \\
&I_{(P)X}^Y \tP^i_Y=\tP^i_X, \qquad
I_{(P)X}^Y \tQ^i_Y=0, &
\plabel{projpr}\nonumber
\eeq
and by similar relations for $I_{(Q)}$ and the inverse multiplets.
The projectors $I_{(Q)}$ and $I_{(P)}$
have also a geometric meaning. Let us consider a non-vanishing
$\chi$ satisfying $\chi^2<1$. 
The latter condition means that by a gauge transformation $\chi$
can be sent to zero. Therefore, the vectors $\chi$ are in one-to-one 
correspondence with boosts and can be thought as boost parameters. 
Then $\chi$ defines a ``boosted'' subgroup \sgchi of \SLC\ which is obtained 
from the canonical embedding of SU(2)
by applying the corresponding boost. The generators of this subgroup in the 
defining representation annihilate the vector $v_\chi=(1-\chi^2)^{-1/2}(1,\chi_a)$.
Then the matrix $I_{(Q)}$ projects the generators of \SLC\ to the generators of
\sgchi and $I_{(P)}$ is a projection to the orthogonal part
(see \cite{AlLiv} for details).

The algebra of the Dirac brackets with the connection \Ref{spcon} takes
the following form
\beq
\{ \tP_X^i(x),\tP_Y^j(y)\}_D&=&0, \nonumber \\
\{ \SA_i^X(x),\tP_Y^j(y)\}_D&=&\delta_i^j I_{(P)Y}^X \delta(x,y),
\plabel{comm} \\
\{ \SA^X_i(x),I_{(P)}^{YZ}(y)\}_D&=&0.
\plabel{chicomm} \nonumber
\eeq
The last relation is important since it shows that the field $\chi$ commutes
both with $\tP$ and $\SA$ and the projectors can be considered as $c$-numbers
with respect to the Dirac algebra.
The commutator of two connections is much more complicated. 
It was derived first in \cite{SAhil}. However, as we show in Appendix \ref{B},
that expression can be further simplified so that 
the final result does not depend on the Immirzi parameter $\im$ and it reads
\be
\{ \SA^X_i(x), \SA^Y_j(y)\}_D = 
\CM_{ij}^{XY}  \delta(x,y),
\plabel{commAAA}
\ee
where $\CM_{ij}^{XY}$ is a $\im$-independent differential operator 
given in \Ref{CMMM}.
Thus, the whole Dirac algebra does not contain the Immirzi parameter.

As usual, the area operator of a two-dimensional surface $\Sigma$ is defined as a
regularization of the classical expression
\be
{\cal S}(\Sigma)=\int_{\Sigma} d^2 \sigma  \sqrt{n_i n_j\, g^{XY}\tP^i_X\tP^j_Y}, 
\plabel{areadef}
\ee
where $n_i$ is the normal to the surface.
The regularization involves a partition $\rho$ of $\Sigma$ into 
small surfaces $\Sigma_n$, $\bigcup_n \Sigma_n=\Sigma$.
Then the regularized area operator is given as a limit of the infinitely fine
partition
\be
{\cal S}=\lim\limits_{\rho \to \infty}\sum\limits_{n}
\sqrt{g(S_n)},
\plabel{areaop}
\ee
where
\beq
g(\Sigma)&=&g^{XY}\tP_X(\Sigma)\tP_Y(\Sigma),   \\
\tP_X(\Sigma)&=&\int_{\Sigma} d^2\sigma  \,
n_i(\sigma) \tP_X^i(\sigma). 
\nonumber
\eeq
The spectrum of the operator \Ref{areaop} follows from the commutation relations 
\Ref{comm} and it is expressed as a combination of two Casimir operators \cite{AV} 
\be
{\cal S}=8\pi \hbar G \sqrt{C(su_{\chi}(2)) - C_1(sl(2,\Cb))},
\plabel{arsp} 
\ee
where we restored the dependence of the Newton's constant.
The subgroup $SU_{\chi}(2)$ depends on the field $\chi$ and was defined
after equation \Ref{projdel}.

\section{Timelike foliation and area operator}

The results reported in the previous section were obtained assuming
that $\chi^2<1$.
However, all algebraic relations up to equation \Ref{areadef} remain valid
also for $\chi^2>1$.
We get a singular situation only if $\chi^2= 1$,
when the described canonical analysis breaks down due to the appearance
of additional constraints.\footnote{It is clear that some constraints must appear because
the condition $\chi^2= 1$ removes one canonical variable.}
What is the physical interpretation of these different cases?
It turns out that they correspond to different causal properties
of the foliation. Indeed, the induced 3-dimensional metric
on the hypersurfaces $t=const$ reads
\be
g g^{ij}=g^{XY}\tP^i_X\tP^j_Y. \plabel{met3}
\ee
Its determinant is 
\be
g\equiv\det g_{ij}=(1-\chi^2)h.
\plabel{detg}
\ee
Since by definition $h>0$ (if the triad is non-degenerate), the cases
$\chi^2<1$ and $\chi^2>1$ can be interpreted
as describing spacelike and timelike foliations, respectively.
It is not surprising that the singular case $\chi^2=1$ corresponds
to a foliation with lightlike slices.

Here we will be interested in the case $\chi^2>1$
of the timelike foliation.
Although in this case the coordinate $t$ can not be identified with time
and the sense of quantization based on the canonical formulation
developed with respect to such spacelike variable is questionable,
we are going to generalize the analysis of
\cite{AV,SAcon} considering the area operator \Ref{areaop}
in the theory with $\chi^2>1$.
At least at the formal level one does not encounter any inconsistencies.

Since the analysis of the commutation relations carried out in \cite{SAcon}
does not depend on the value of $\chi$, we have again a unique
spacetime connection given by \Ref{spcon}, which leads to
a diagonal operator ${\cal S}$.
As in \cite{AV}, using the commutation law (\ref{comm}), one obtains
\begin{equation}
{\cal S} \sim\hbar
\sqrt{-I_{(P)}^{XY}T_XT_Y},\plabel{au}
\end{equation}
where $T_X$ are $sl(2,\Cb)$ generators.
Thus, the spectrum is completely determined by the properties of the projectors. 
As we mentioned in the previous section, as soon as $\chi^2<1$, 
$\IQ$ projects to the boosted subgroup \sgchi of \SLC.
Therefore, in this case the operator
\be
C=I_{(Q)}^{XY}T_X T_Y  \plabel{Cas}
\ee
coincides with the Casimir operator of \SUU, which together with the first relation
in \Ref{projdel} leads to the spectrum \Ref{arsp}.

Now we derive an analogous statement for the case $\chi^2>1$.
Let us introduce the generators
\begin{equation}
q_a:=\frac{1}{\sqrt{\chi^2-1}}\left(\delta_{ab}-
\frac{1-\sqrt{\chi^2-1}}{\chi^2}\chi_a\chi_b\right)
\Et_i^b \tQ^i_X T^X,
\plabel{qa}
\end{equation}
and the metric which will play the role of the Killing 
form\footnote{The expression for the Killing form (\ref{kilf})
differs essentially from that implicitly used in \cite{AV}, $k^{ab}=\delta^{ab}$, 
for $\chi^2<1$. In fact, one can describe all $\chi^2 \ne 1$ in a uniform manner. 
Let us take the generators as
$$
q_a=\sign(1-\chi^2)\left(\delta_{ab}-
2\frac{{\chi_a\chi_b}}{\chi^2}\right)
\Et_i^b \tQ^i_X T^X.
$$
Then
\beq
&  I_{(Q)}^{XY}T_XT_Y=-k^{ab}q_aq_b, & \nonumber \\
& [q_a,q_b]=-|\det k|^{-1/2}\eps_{abd}k^{dc}q_c , & \nonumber
\eeq
where 
$$
k^{ab}=\frac{\delta^{ab}-\chi^a\chi^b}{1-\chi^2}.
$$
Thus, the structure constants are
$f_{ab}^c=-|\det k|^{-1/2}\eps_{abd}k^{dc}$.
One can check that $k^{ab}$ is the Killing form of the algebra
generated by $q_a$ because
$$
f_{ac}^d f_{bd}^c= -2k_{ab},
$$
where $k_{ab}$ is the inverse of $k^{ab}$.
The signature of the Killing form is defined by the value of $\chi$.
For $\chi^2<1$ it is $(+,+,+)$, whereas for for $\chi^2>1$ it changes to $(+,+,-)$.
}
\be
k^{ab}:=\delta^{ab}-\frac{2\chi^a\chi^b}{\chi^2}.
\plabel{kilf}
\ee
One can check directly that
\beq
&  I_{(Q)}^{XY}T_XT_Y=k^{ab}q_aq_b, & \\
& [q_a,q_b]=-\eps_{abd}k^{dc}q_c , & \plabel{qcom}
\eeq
so that $f_{ab}^c=-\eps_{abd}k^{dc}$ are the structure constants of the algebra 
generated by $q_a$ and $k_{ab}=\hf f_{ac}^d f_{bd}^c$ is indeed its Killing form.
The signature of $k^{ab}$ is $(+,+,-)$. Therefore, we conclude that the corresponding
algebra is $sl(2,\Rb)$ and the operator (\ref{Cas}) is now its Casimir operator.
As a result, the area spectrum is
\begin{equation}
{\cal S}=8\pi \hbar G  \sqrt{C(sl_{\chi}(2,\Rb))-C_1(sl(2,\Cb))}
\,.\plabel{areaspec}
\end{equation}
Hence, changing the foliation from spacelike to timelike corresponds to
the replacement of the \SUU\ subgroup of the Lorentz gauge group by \SLR.
This is quite natural because the subgroup is always associated with
the symmetry group in the tangent space of the slices.

The explicit form of the spectrum can be obtained taking into account
the values of the Casimir operators for irreducible representations.
Let us restrict ourselves to the principal series of
representations of the Lorentz group.\footnote{These are representations
appearing in the decomposition of a square integrable function on \SLC.} 
They are labeled by two numbers $(n,\rho)$, where $n\in \Nb/2,\ \rho\in \Rb$.
In a decomposition of this representation with respect to the subgroup \SLR\
there appear the principle continuous
series of unitary representations of SL(2,\Rb)
and a finite number of discrete series representations. The latter are
labeled by $0\le j < n,\ n-j\in \Nb$. Thus, we arrive at two possibilities:
\beq
\CS &\sim & \hbar\sqrt{-\(\frac{1}{4}+s^2\)-n^2+\rho^2+1}, \plabel{cont} \\
\CS &\sim & \hbar\sqrt{j(j+1)-n^2+\rho^2+1}.  \plabel{discr}
\eeq

They exhaust the possible forms of the spectrum of the area operator. 
However, the actual spectrum might be given by only a subset of the representations
appearing in \Ref{cont} and \Ref{discr}. This subset should be determined by the 
construction of the Hilbert space of CLQG. Unfortunately, it is still lacking
due to the difficulty to impose the second class constraints at the quantum level. 
Nevertheless, in the next section we show that the result found here contains 
the spectrum coming from the Lorentzian spin foam model of \cite{sfLor2}.

\section{States and area spectrum: canonical versus spin foam approach}

\subsection{Enlarged Hilbert space: Projected spin networks}

In contrast to \slqg, the construction of the kinematical Hilbert space
(the space on which one should then impose the quantum constraints
corresponding to the first class constraints of the classical theory)
in the framework of CLQG is a two-step procedure.
In the beginning we construct a Hilbert space of functionals of arbitrary
Lorentz connections. It is an enlarged space because the connections
appearing in our formalism are not arbitrary but subject to the second
class constraints \Ref{contA}. Therefore, the second step is to impose these
constraints at the level of the Hilbert space.

This is a non-trivial problem because in the loop approach the elements of the
Hilbert space are (multi)loop states or spin network states.
They are characterized not by a functional dependence on the connection, which
is similar for all states and described through holonomies, but by irreducible
representations assigned to the loops or to the edges of a graph. Therefore,
the implementation of the second class constraints means a restriction on possible
representations. It is not evident {\it a priori}
how the condition \Ref{contA} on the form of connections
can be encoded in such a restriction and that this is possible at all.
We leave the solution of this problem for future research, but in the next subsection
we will see that a certain restriction is in accordance with the spin foam models.
Here we only provide a description of the enlarged Hilbert space.

In fact, the elements of the enlarged Hilbert space are gauge invariant
functionals of both the Lorentz connection and the field $\chi$.
This can be justified by the fact that
the connection $\SA$, which will be used in the definition of the holonomies, commutes 
with $\chi$
(see \Ref{comm}), so that the two fields can be simultaneously considered
as configuration variables. Since we are following the loop approach,
the Hilbert space structure should be similar to what one has in \slqg.
In particular, we expect that the basis is realized by spin network like states
and the scalar product must be Lorentz and diffeomorphism invariant.
Such a Hilbert space was constructed in \cite{psn} and we review here
the main results. The construction in the case of a temporal foliation differs
only in small details. Therefore, we present it in a general form.  

The necessary Hilbert space structure can be induced from the space of
the so-called projected cylindrical functions. Let us consider an oriented graph $\Gamma$
with $E$ edges and $V$ vertices. With each edge $\gamma_k$
we associate a holonomy of the Lorentz connection $\SA$:
$U_{\gamma_k}[\SA]=\CP \exp \(\int_{\gamma_k} \SA \)$ which gives an element
$g_k$ of the Lorentz group. Besides, with each vertex $v_r$ we associate
an element $x\in X\equiv{\rm SL}(2,\Cb)/H$, where $H={\rm SU}(2)$ or \SLR\ depending
on whether the foliation is spacelike or timelike. 
It is defined by the field $\chi$ as follows \cite{AlLiv}:
\be
x(\chi)=\( {1\over\sqrt{|1-\chi^2|}},{\chi^a\over\sqrt{|1-\chi^2|}}\),
\plabel{xchi}
\ee
where we use identification of $X$ with one of the hyperboloids
in Minkowski space. Also pick a complex valued function
$f(g_1,\dots,g_E;x_1,\dots,x_V)$ on $[\SLC]^E\otimes [X]^V$ which satisfies the following
invariance property
\be
f(\gl_{t(1)} g_1 \gl_{s(1)}^{-1},\dots,\gl_{t(E)}g_E \gl_{s(E)}^{-1};
 \gl_1\cdot x_1,\dots, \gl_V\cdot x_V)=f(g_1,\dots,g_E;x_1,\dots,x_V),
\plabel{invprop}
\ee
where $t(k)$ and $s(k)$ denote, respectively,
the target and the source vertex of the $k$th edge of the graph $\Gamma$,
$\gl_r \in \SLC$ and its action on $x_r$ coincides with the usual Lorentz transformation.
Then the projected cylindrical function is defined as
\be
F_{\Gamma,f}[\SA,\chi]=
f\(U_{\gamma_1}[\SA],\dots,U_{\gamma_E}[\SA];x(\chi(v_1)),\dots,x(\chi(v_V))\).
\plabel{pcf}
\ee
Due to the property \Ref{invprop}, it is invariant under local Lorentz
transformations. The set of all projected cylindrical functions is dense
in the space of all smooth gauge invariant functionals of $\SA$
and $\chi$.

It is easy to define a gauge invariant scalar product on the cylindrical functions.
It is given by
\be
\langle F_{\Gamma,f}|F_{\Gamma',f'}\rangle =
\int_{[\SLC]^E} \prod\limits_{k=1}^E
d g_k \, \overline{ f (g_1,\dots,g_E;x_1,\dots,x_V )}
f' (g_1,\dots,g_E;x_1,\dots,x_V ),
\plabel{scpr}
\ee
where we imply the usual extension of the functions
to the unified graph $\Gamma\cup\Gamma'$.
Note that the integration over the variables $x_r$, which would correspond to
the integration over the field $\chi$ in the path integral approach,
is missing. These variables are fixed and can be chosen arbitrary.
Due to the invariance of the Haar measure, the scalar product does not depend on
this choice. One can say that the fixing of $x_r$ corresponds to the necessity
to fix a gauge in the path integral approach.
The enlarged Hilbert space $\CH_0$ of CLQG is obtained as
the completion of the space of the projected cylindrical functions
in the norm induced by the bilinear form \Ref{scpr}.

An important set of states in $\CH_0$ is realized
by the so-called {\em projected spin networks} \cite{psn}.             
To introduce these objects, let us consider a graph with the following `coloring'.
With each edge $\gamma_k$ we associate an irreducible representation
of \SLC\ from the principle series, $\lambda_k=(\rho_k,n_k)$,
and two representations of $H$, $j_{t(k)}$ and $j_{s(k)}$, appearing in
the decomposition of $\lambda_k$ on the subgroup $H$ (for the continuous series
of \SLR, $j=is-\hf$). The first representation is attached to the final point
of the edge, and the second one corresponds to its beginning.
With each vertex $v_r$ we associate an intertwiner $N_r$ between the 
representations of $H$ attached to the ends meeting at this vertex.

We are going to construct a cylindrical function according to this coloring.
For this we take holonomies of the Lorentz connection $\SA$ in the representations
$\lambda_k$ along the edges, project them at the ends on the representations of $H$ and
contract the resulting objects with the intertwiners $N_r$ at the vertices.
To make this procedure clear, one should explain the meaning of the projection
on the representations of the subgroup and the way how it works.

The origin of this projection and the possibility to do it in a Lorentz covariant
way can be traced back to the presence of the field $\chi$ \cite{SAhil}.
As we explained above, it defines a boosted subgroup $H_\chi$ of \SLC. 
This subgroup is a stationary group of
the vector $x(\chi)$ \Ref{xchi}. Any representation of \SLC\ can be decomposed
into a direct sum (integral) of irreducible representations of $H_\chi$
\be
\CH_{\SLC}^{\lambda}=\mathop{\bigoplus}\limits \CH_{H_\chi}^j.
\plabel{decrepr}
\ee
The orthogonal projectors on each component of the decomposition are called
{\it projective operators} and can be written explicitly as follows
\be
\Ppr{\lambda,j}{\chi}=d_j\int_{H_\chi} dh\, \overline{\CC^{j}(h)}D^{(\lambda)}_{\SLC}(h).
\plabel{Ppr}
\ee
Here for finite dimensional representations of the subgroup, 
$d_j=2j+1$ is the dimension of the representation $j$,
$\CC^{j}(h)=\tr[ D^{(j)}_{\rm SU(2)}(h)]$ is its character, and
$D^{(\lambda)}_{\SLC}$ is the representation matrix. For infinite dimensional 
representations, $d_j$ is the spectral measure and $\CC^{j}(h)$ should be 
viewed as a distribution.
Since the holonomy $U_{\gamma}^{(\lambda)}[\SA]$ in a representation $\lambda$
can be considered as an element of
$\CH_{\SLC}^{\lambda}\otimes \CH_{\SLC}^{\lambda}$, one can introduce
{\it projected Wilson lines} by applying the projective operators \Ref{Ppr}
from the two ends of the line \cite{SAhil}
\be
U^{(\lambda;j_1,j_2)}_{\gamma}[\SA,\chi]=
\Ppr{\lambda,j_1}{\chi(v_t)} U_{\gamma}^{(\lambda)}[\SA]
\Ppr{\lambda,j_2}{\chi(v_s)}.
\plabel{WLpe}
\ee
This procedure gives an element of $\CH_{H_\chi}^{j_1}\otimes \CH_{H_\chi}^{j_2}$.

One could think that the projection on a subgroup spoils the covariance of
the Wilson lines.
However, the dependence on $\chi$ restores the covariance under the local
Lorentz transformations. Indeed, it is easy to check from the explicit form \Ref{Ppr}
that the projective operators transform homogeneously
\be
\Ppr{\lambda,j}{{}^{\gl}\chi}=
D^{(\lambda)}_{\SLC}(\gl)\Ppr{\lambda,j}{\chi} D^{(\lambda)}_{\SLC}(\gl^{-1}),
\plabel{trPpr}
\ee
where ${}^{\gl}\chi$ is the Lorentz transform of $\chi$ by $\gl\in \SLC$.
This property immediately gives the usual transformation law for the projected
Wilson lines as it would be a simple holonomy of a Lorentz connection
\be
U^{(\lambda;j_1,j_2)}_{\gamma}[{}^{\gl}\SA,{}^{\gl}\chi]=
D^{(\lambda)}_{\SLC}(\gl_t)U^{(\lambda;j_1,j_2)}_{\gamma}[\SA,\chi]
D^{(\lambda)}_{\SLC}(\gl_s^{-1}).
\plabel{trWL}
\ee

Taking into account all these definitions, a projected spin network can be written as
a scalar product
\be
\Psi_S[\SA,\chi]=
\mathop{\bigotimes}\limits_{k=1}^E
U^{(\lambda_k;j_{t(k)},j_{s(k)})}_{\gamma_k}[\SA,\chi]
 \cdot \mathop{\bigotimes}\limits_{r=1}^V  \iota_{(\chi(v_r))}\( N_r\),
\plabel{projspin}
\ee
where $S=(\Gamma,\vec \lambda, \vec j_t, \vec j_s, \vec N)$ is a collection of
the graph and its coloring. The symbol $\iota_{(\chi(v))}(N)$ denotes an embedding
of the intertwiner $N$, which is an element of
$\mathop{\otimes}_{\gamma_k \ni v} \CH_{H_\chi}^{j_k}$, into the space
$\mathop{\otimes}_{\gamma_k \ni v} \CH_{\SLC}^{\lambda_k}$.
This embedding is necessary to ensure the Lorentz invariance of the spin networks.
The embedding depends on $\chi$. Therefore, similarly to the projected Wilson lines,
the embedded intertwiner transforms in a covariant way. Together with \Ref{trWL}
it is enough for $\Psi_S$ to be gauge invariant. It is evident that the projected
spin networks with different coloring are orthonormal with respect to the scalar product
\Ref{scpr}.\footnote{In \cite{psn} it was also argued that the projected spin networks
form a complete set of states, {\it i.e.}, an orthonormal basis in $\CH_0$. 
However, we are not aware of any proof of this statement. 
Therefore, we avoid to view them as a basis.}

Note that in the case of spacelike foliation the projection used to define
the projected spin networks effectively reduces
all infinite dimensional representations to finite dimensional subspaces. 
Therefore, the scalar product in \Ref{projspin} is well defined. 
It gives traces only over finite dimensional representation spaces of SU(2).
This fact is quite helpful because there was an attempt to define the usual
spin networks for non-compact gauge groups \cite{FrLiv,noncomp}, 
which showed that it is a quite non-trivial problem. 
In our approach the solution comes from the use of the 
projection to a subgroup and avoids any mathematical complications.
Of course, for a timelike foliation, all difficulties remain since the representations
to project to are still infinite dimensional. However, since in this case 
we do not expect to obtain a meaningful quantum theory, such problems are not
crucial.

The usual (non-projected) spin networks can be obtained as a
sum (or integral) of the projected spin networks over all representations 
of the subgroup associated with the edges. They form only a small subset of all states
in our approach. Moreover, they do not even belong to the Hilbert space $\CH_0$ 
since they are not normalizable. 
This shows that considering spin networks constructed only from
the connection, one misses an important information. When this information
(dependence on $\chi$) is restored, many problems are solved automatically.

But the main advantage of the projected spin networks is that they are eigenstates
of the area operator \Ref{areaop}.
Although the action of the area operator on a Wilson line of the Lorentz connection
$\SA$ is expressed through the Casimir operators, it is not yet diagonal because
of the Casimir operator of $H_\chi$. This operator takes different values on
different $\CH_{H_\chi}^{j}$, subspaces of the representation space of the Lorentz group.
Therefore, it becomes diagonal only after a projection on one of these subspaces.
The projected spin networks just accomplish this requirement.
Note, however, that the projection is done only at the vertices. Hence, the 
states \Ref{projspin} are eigenstates of the area operators of only those surfaces, which intersect
the graph underlying the spin network near vertices. In \cite{SAhil} another construction
was suggested where the projection is done at every point of edges, so that
the resulting spin networks are eigenstates of all areas. But its relation to the spin foam
models is more subtle. Instead, we will concentrate here on states where the projections 
are done only at the vertices.

\subsection{Comparison with spin foam models}

Although we cannot rigorously decide which representations labeling the 
projected spin networks survive after imposing the second class constraints,
one can find their subset which allows to recover all states arising 
from the Lorentzian spin foam models \cite{BC,sfLor1,sfLor2}.
First of all, let us briefly describe how a spin foam induces 
spin networks states on a foliation of spacetime.
This has been shown in the recent work \cite{Maran3}, so we just recall
the main steps of the construction. 

Let us start with theories with a BF-type action. 
A spin foam amplitude can be considered as a discretization of the 
path integral and arises upon a triangulation of spacetime.
It contains an integral over the gauge group for 
each {edge} (1-codimensional simplices) of the triangulation. 
The integrand is the direct product of group elements in the 
representations associated to the bones (2-codimensional simplex) incident to 
the edge. Performing such an {\it edge integral} gives the product of two 
intertwiners corresponding to the two sides of the edge. If one introduces a 
slicing of the triangulation and performs the integrals for edges lying in one 
simplicial level hypersurface $\Sigma_k$ only, then one obtains two spin networks 
$\psi_k^+$ and $\psi_k^-$ associated to the two sides of the slice. 
They are defined on the graph dual to the triangulation of $\Sigma_k$, {\it i.e.}, 
their vertices and links correspond to the edges and bones, respectively, 
belonging to $\Sigma_k$. The group element assigned to a link of the graph 
is the product of the group elements assigned to the edges lying between $\Sigma_k$ 
and the neighboring level $\Sigma_{k\pm 1}$ and incident to the dual bone. 
It is clear that the spin networks $\psi_k^\pm$ differ only by these group elements,
whereas all their labels coincide (up to conjugation). In contrast, the spin networks
associated with different slices can differ essentially from each other what corresponds
to insertion of an interaction between the two slices.
The spin foam partition function can be recovered by (i) taking a certain
scalar product of the spin networks $\psi_k^+$ and $\psi_{k+1}^-$, which consists
essentially in evaluation of the group integrals for edges lying between $\Sigma_k$ 
and $\Sigma_{k+1}$, (ii) by summing (or integrating) over the representations
associated to the spin networks, and finally (iii) by summing over slices $\Sigma_k$.

To obtain a spin foam model of general relativity, one should impose additional
constraints on the allowed representations and intertwiners which decorate the faces 
and edges of the dual two-simplex of the triangulation of the manifold \cite{BC}.
First, the representations are restricted to the so called simple representations 
of $\SLC$, which are characterized by vanishing of the second Casimir operator 
$C_2(sl(2,\Cb))=n\rho=0$. Thus, one always has either $n=0$ or $\rho=0$.
Secondly, the intertwiners are given by the so called Barett-Crane intertwiner
to be described below.
The resulting spin networks are called simple spin networks 
\cite{BFhigher,simplespin}. 

To make the identification between simple and projected spin networks as 
explicit as possible, we use the fact that any simple representation $\lambda$ can 
be realized on functions $f^{(\lambda)}(x)$ on the homogeneous space $X={\rm SL}(2,\Cb)/H$
(see, for example, \cite{barack,ruhl}).
The choice of $H$ depends on which series of representations, continuous or discrete, 
is considered. The series $(n,0)$ appears only for $H=\SLR$, whereas the series $(0,\rho)$
can be obtained for both maximal subgroups.
Correspondingly, there are two spin foam models based either on SU(2) \cite{sfLor1}
or \SLR\ \cite{sfLor2}. In the former only the simple representations from the
continuous series can label the faces, whereas the latter admits both types of 
representations. It is clear that in our case the relevant choice is $H=\SLR$ 
so that both series will appear.

Thus, let $\lambda$ be a simple representation and $\{f_p^{(\lambda)}\}$ 
be an orthonormal basis in $\CH_{\SLC}^{\lambda}$. Then the matrix elements of 
$g\in \SLC$ in the representation $\lambda$ can be written as an integral over the
homogeneous space
\be 
\( D^{(\lambda)}_{\SLC}(g)\)_{pq}=
\int_X dx \, f_p^{(\lambda)}(g\cdot x)\overline{f_q^{(\lambda)}(x)}.
\plabel{matrel}
\ee
Besides, in this basis the Barrett-Crane intertwiner, 
which is the only one allowed in the simple spin networks, 
can be represented at a $l$-valent vertex with $i$ incoming and $l-i$ outgoing links as 
follows \cite{BFhigher,simplespin} 
\be 
N_{p_1\dots p_i,q_{i+1}\dots q_l}^{(BC)\,\lambda_{1}\dots\lambda_l}=
\int_X dx \, \overline{f^{(\lambda_{1})}_{p_1}(x)}\cdots 
\overline{f^{(\lambda_{i})}_{p_i}(x)}
f_{q_{i+1}}^{(\lambda_{i+1})}(x)\cdots f_{q_l}^{(\lambda_l)}(x).  
\ee
A simple spin network is given by coupling of these intertwiners at vertices with
the matrix elements \Ref{matrel} associated with links of the underlying graph.
Performing the summation, one finds
\be
\Phi_{(\Gamma,\vec\lambda,\vec N^{(BC)})}[\vec g\,]=
\prod\limits_{r=1}^V\int_X dx_r \,\,
\prod\limits_{k=1}^E K^{(\lambda_k)}\(x_{t(k)},g_k\cdot x_{s(k)}\),
\plabel{simsn}
\ee
where the kernel is defined as  
\be 
K^{(\lambda)}(x,y)\equiv 
\sum_{p}\overline{f_p^{(\lambda)}(x)} f^{(\lambda)}_p(y)=K^{(\lambda)}(\theta(x,y)). 
\ee
Here, as usual, $x_{s(k)}$ and $x_{t(k)}$ are the integration variables 
at the source and target vertices of the $k$-th link 
and $\theta(x,y)$ is the hyperbolic distance between $x$ and $y$. 

For the case of spacelike foliation when $H={\rm SU}(2)$, in \cite{psn,AlLiv}
it was already shown that the simple spin networks \Ref{simsn} are identical to the 
(integrated with respect to $x_r$) projected spin networks \Ref{projspin} with 
$\lambda_k=(0,\rho_k)$ and $j_{t(k)}=j_{s(k)}=0$. It means that the projection 
should be always done to the trivial representation of the subgroup.
As we will see now, a similar result is valid also for the timelike case
with the only difference that the representations of type $(n,0)$
are also admissible.

Let us identify the ingredients used in the construction of the simple spin networks 
with the basic elements of the projected spin networks. 
First of all, it is clear that, considering $x_0$ as a parameter, 
the kernel $K^{(\lambda)}(x_0,x)$ can be viewed
as a vector in $\CH_{\SLC}^{\lambda}$, which is invariant with respect to the subgroup 
$g_0 H g_0^{-1}$ where $g_0$ is a representative of $x_0$ in \SLC.
In other words, it is invariant under the subgroup boosted with the parameter defined by $x_0$.
The functions $\overline{f_p^{(\lambda)}(x_0)}$ are the components of this vector in 
our basis and the combination
\be
\(\Ppr{\lambda,0}{\chi}\)_{p p'}=
f_{p}^{(\lambda)}(x(\chi))\overline{f_{p'}^{(\lambda)}(x(\chi))}
\plabel{projinv}
\ee 
can be identified with the matrix elements of the projector \Ref{Ppr}
to the singlet representation $j=0$ of the subgroup $H_{\chi}$.\footnote{The existence
of the singlet component in the decomposition with respect to a maximal subgroup 
is a characteristic property of simple representations.}
The only complication arising for $H=\SLR$ is that the invariant vector is not 
normalizable and the projector should be understood as a distribution.
The corresponding projected Wilson line is
\be
\(U^{(\lambda;0,0)}_{\gamma}[\SA,\chi]\)_{pq}=
f_{p}^{(\lambda)}(x_{t}) K^{(\lambda)}\(x_{t},U_{\gamma}[\SA]\cdot x_{s}\) 
\overline{f_{q}^{(\lambda)}(x_{s})},
\plabel{projWLx}
\ee
where $x_r=x(\chi(v_r))$ is defined by \Ref{xchi}.
Similarly, the tensor
\be
\({\iota_{(\chi)}\({\bf 1}\)}\)_{p_1\dots p_i,q_{i+1}\dots q_l}=
\overline{f^{(\lambda_{1})}_{p_1}(x(\chi))}\cdots 
\overline{f^{(\lambda_{i})}_{p_i}(x(\chi))}
f_{q_{i+1}}^{(\lambda_{i+1})}(x(\chi))\cdots f_{q_l}^{(\lambda_l)}(x(\chi))
\plabel{BCtens}
\ee
is the embedding of the trivial intertwiner between $l$ singlet representations
of the boosted $H$ into the tensor product of simple representations $\{\lambda_k\}$ of \SLC. 
Taking this into account, as well as the defining relations 
\Ref{simsn} and \Ref{projspin}, one finds
\be
\Phi_{(\Gamma,\vec\lambda,\vec N^{(BC)})}\[\vec U_{\gamma_k}[\SA]\]=
\prod\limits_{r=1}^V\int_X dx_r \,\,
\Psi_{(\Gamma,\vec\lambda,0,0,{\bf 1})}[\SA,\chi(x)],
\plabel{relsnsn}
\ee
where $\vec \lambda$ is a set of simple representations and 
we neglected the factor $\prod_{k=1}^E K^{(\lambda_k)}(0)$ which is equal to 1
for the compact subgroup $H$ and is infinite in the non-compact case.
Thus, we conclude that all boundary states induced by the
Lorentzian spin foam models are described by the projected spin networks
of CLQG, where all representations labeling edges are simple ones and 
all projections are done to the singlet representation of the $\chi$-dependent subgroup. 
In particular, for a timelike foliation the projected spin networks reproduce
all boundary states of the $\SLC/\SLR$ model of \cite{sfLor2}.

Since the Casimir operator of \SLR\ vanishes on the singlet representation,
on the states appearing in \Ref{relsnsn} the area spectrum \Ref{areaspec} reduces to 
\be
\CS \sim \hbar\sqrt{1+\rho^2} \qquad {\rm or} \qquad
\CS \sim \hbar\sqrt{1-n^2}
\plabel{redarsp}
\ee
for the continuous and discrete series of simple representations, respectively.
This result coincides with the prediction of the spin foam model,
where the area operator is represented simply as 
$\sqrt{-C_1(sl(2,\Cb))}$ from the very beginning \cite{BC}. 
We observe that the first spectrum in \Ref{redarsp} is real, whereas the second
is imaginary. This agrees with the fact that a timelike 3-dimensional slice may contain
both spacelike and timelike surfaces and confirms the expectation that their
areas are continuous and discrete, correspondingly.

\section{Discussion}

In this paper we presented the covariant loop quantization of a canonical
formulation based on a timelike foliation of spacetime.
We want to emphasize that this quantization is very formal 
and should not be considered as a model for quantum gravity
because of a wrong (spacelike) direction of the evolution in such a formulation.
The aim of the construction was just to show that it produces exactly
the same structures which arise in some Lorentzian spin foam models 
of quantum gravity. We hope that this can help to understand better
the relation between the canonical and the spin foam approaches.

We showed that the projected spin networks arising in the covariant quantization
projected on the singlet representation of \SLR\ reproduce the simple spin 
networks of the $\SLC/\SLR$ spin foam model. The area spectrum evaluated on 
these states also perfectly agrees with the one predicted from the spin foam approach.
In particular, it implies that the area spectrum of spacelike surfaces is continuous 
and that of timelike surfaces is discrete. This seems to be a very general observation
(which by the way is not true in SLQG) because it was found also in other two models
of quantum gravity.  

First, it appears as a result for the length spectrum in $2+1$ gravity 
both in the canonical loop quantization \cite{FLR} and in the spin foam approach. 
In that situation the spectrum depends only on the Casimir operator of
the full gauge group, which is \SLR, and no contribution from a subgroup appears.
As in the 4-dimensional case, the continuous series of representations
is associated with spacelike lines and the discrete series corresponds to timelike lines.

The second place where one arrives at a similar conclusion is the 't Hooft model
of $2+1$ quantum gravity coupled to a point-like particle \cite{thooft1,thooft2}.
A careful canonical analysis of this system shows that the spatial position of
the particle has a continuous spectrum, whereas temporal positions are quantized 
\cite{thooftst}.
Note that in \cite{zoltan} a relation between the 't Hooft model and the algebra of loop
variables for the case of several gravitating particles was established, whereas in 
\cite{frlo} the possibility to connect the 't Hooft model to the spin foam quantization
was pointed out.

Thus, the qualitative result --- ``space is continuous, time is discrete" --- seems 
to be a general feature of diffeomorphism invariant theories.
It is very plausible that CLQG is in agreement with this picture.
Of course, it may cause difficulties in deriving the black hole entropy
since the standard derivation of SLQG, which relies essentially on the discreteness
of the area and its dependence of the Immirzi parameter, does not work anymore. 
However, as it was shown recently in \cite{SAbh}, the continuity of the area spectrum 
is not a serious obstacle on this way.  

Although we identified the physical sense of the two series of representations
corresponding to the states described by the simple spin networks, 
the canonical theory contains much more states. In particular, if 
the projection is done not on the singlet representation, {\it i.e.,} $j\ne 0$,
the corresponding area operator has a more complicated form with a contribution
from the Casimir of the subgroup. What is the meaning of the other branches
of the spectrum?

One possibility would be that the other states are simply excluded from the kinematical 
Hilbert space, for example, by imposing the second class constraints as it was 
discussed in the beginning of section 4.1.
This is supported by the comparison with the spin foam models which know nothing about
these additional states.
However, at the moment the spin foam approach is not 
justified rigorously and it is not guaranteed that the true model of quantum gravity 
will not differ at least in some details.
Thus, one cannot be sure that the presence of the Casimir operator of the subgroup
in the area spectrum is meaningless. 

Its appearance can be traced back to the existence of the second class constraints. 
They change the commutation relations since the Poisson bracket must be 
replaced with the Dirac bracket as described in section 2. 
There are two sets of such constraints. On the other hand, 
as it was shown in \cite{AlLiv}, in the spin foam models
only one set, the so called simplicity constraint,
is taken into account. But it is the second set which is
the most complicated one and leads to nontrivial commutation relations.
For this reason it seems that the present spin foam models should be complemented with
an additional ingredient to be consistent with the canonical approach.

An attempt to address this problem constructing a rigorous canonical analysis
of Plebanski formulation of general relativity was done in \cite{canBF}.
The construction again produces some second class constraints which 
can be viewed, of course, as our constraints \Ref{phi}
and \Ref{psi} written in new variables. The resulting Dirac brackets of \cite{canBF}
are even more complicated than in the covariant formulation discussed here and 
it is not clear how to take them into account in the spin foam quantization. 
We hope that our work may shed some light on this issue.

Note that there is another feature distinguishing between the structures coming
from the covariant loop approach and from the spin foam models.
Namely, the projected spin networks of CLQG are functionals of not only the connection, but
also of the field $\chi$. In the simple spin networks arising as boundary states 
of spin foams this dependence is removed by integrating with respect to variables $x_r$
living at the vertices (see \Ref{relsnsn}). However, this integration might kill some
important information. Indeed, the degrees of freedom represented by the field $\chi$ 
are decoupled from the other fields and carry information additional 
to that contained in the connection.
In the projected spin networks this is reflected in the assignment 
of representations: whereas the representations of the gauge group
characterize the causal structure, one may think that the representations 
of the subgroup and its choice itself
are associated to a particular foliation determined by $\chi$.
It seems that the extra integration performed in the spin foam models in some sense smears
over all foliations.

In conclusion we would like to stress that the analysis of this paper shows that
the covariant canonical formulation \Ref{dact} allows to treat an arbitrary 
foliation in a uniform manner, similarly as it is done in the $2+1$ case \cite{FLR}. 
For this it is enough to indicate 
whether $\chi^2$ less, equal or larger than $1$.\footnote{The case $\chi^2=1$
may require a special attention.} In principle,
it is even possible to consider all these cases simultaneously
assuming that $\chi^2$ can vary as we wish along the slices. 
This feature might open the possibility to describe fluctuations of 
the causal structure in the framework of the canonical approach.

\appendix

\section{Matrix algebra}
\label{A}

In this appendix we present several useful relations.
First, let us consider the matrices introduced in \Ref{relPQ}
and \Ref{matR}.
Their inverse are
\be
(\Pi^{-1})^Y_X=-\Pi^Y_X, \qquad (R^{-1})^{XY}=
\frac{g^{XY}+\frac{1}{\beta} \Pi^{XY}}{1+\frac{1}{\beta^2}}.
\plabel{inverse}
\ee
Due to these relations $\Pi^X_Y$, $R^{X}_{Y}$ and their inverse commute
with each other.
Furthermore, they commute with the structure constants
in the following sense:
\be
 f^{XYZ'}\Pi_{Z'}^Z=f^{XY'Z}\Pi_{Y'}^Y, \qquad
 f^{XYZ'} R_{Z'}^Z=f^{XY'Z} R_{Y'}^Y. \plabel{com-fr}
\ee

The structure constants are given by the following table
\be
\begin{array}{ccc}
f_{a_1 a_2}^{a_3}=0,&
f_{a_1 b_2}^{a_3}=-\eps^{a_1 b_2 a_3},&
f_{b_1 b_2}^{a_3}=0, \\
f_{b_1 b_2}^{b_3}=-\eps^{b_1 b_2 b_3},&
f_{a_1 b_2}^{b_3}=0,&
f_{a_1 a_2}^{b_3}=\eps^{a_1 a_2 b_3}.
\end{array}      \plabel{algHP}
\ee
Here we split the 6-dimensional index $X$ into a pair of 3-demensional
indices, $X=(a,b)$, so that the indices $a$ correspond
to the Lorentz boosts, whereas the indices $b$ label the SO(3) subgroup.
The contraction of two structure constants can be represented through the
Killing form and the matrix $\Pi$ as follows
\be
f^T_{XY} f^W_{TZ}=-g_{XZ}\delta^W_Y+g_{YZ}\delta^W_X+
\Pi_{XZ}\Pi^W_Y-\Pi_{YZ}\Pi^W_X.  \plabel{ff}
\ee

\section{Commutator of two connections}
\label{B}

In \cite{SAhil} the following expression for the Dirac bracket of the two shifted
connections \Ref{spcon} was obtained
\be
\{ \SA^X_i(x), \SA^Y_j(y)\}_D=
\frac{1}{2\left(1+\frac{1}{\beta^2}\right)} R^X_Z R^Y_W \CK_{ij}^{ZW} \delta(x,y),
\plabel{commAA}
\ee
where $\CK_{ij}^{ZW}$ is a $\im$-independent linear differential operator.

From \Ref{commAA} several important properties of this
commutator can be derived without using the explicit form of $\CK_{ij}^{ZW}$.
Let us redefine the connection \Ref{spcon} by a term proportional to the second
class constraints. Such a shift will not modify the Dirac brackets. We choose
the following redefinition \cite{AlLiv}
\beq
\tSA_i^X &=&\SA_i^X- \frac12 R^X_{Y}
\left(\Qt_l^Y(\Qt\Qt)_{ik}-\frac12 \Qt_i^Y(\Qt\Qt)_{lk}\right)\psi^{lk}
\nonumber \\
&=& \left(1+\frac{1}{\im^2}\right) I_{(P)Y}^X (R^{-1})^Y_Z A_i^Z
+ R^X_Y \Gamma_i^Y,
\plabel{spconnew}
\eeq
where
\be
\Gamma_i^X = \frac12 f^{W}_{YZ}I_{(Q)}^{XY} \Qt_i^Z \p_l \tQ^l_W
+\frac12 f^{ZW}_Y\left((\Qt\Qt)_{ij} I_{(Q)}^{XY}
+\Qt_j^X\Qt_i^Y -\Qt_i^X\Qt_j^Y \right) \tQ^l_Z \p_l \tQ^j_W
\plabel{gam}
\ee
is related to the \SLC\ connection compatible with the metric induced on 
the 3-dimensional slices (see Appendix C).
It is clear that the connection \Ref{spconnew} satisfies the constraints:
\be
I_{(Q)Y}^X\tSA_i^Y=\Gamma_i^X(\tQ),
\plabel{contA}
\ee
which can be thought as a realization of the second class
constraints.\footnote{More precisely, due to the projection
the equation \Ref{contA} contains only 9 non-trivial relations. 
But only six of them are due to the second class constraints.
The other three relations appear because three components of the
initial canonical connection are missing in $\SA_i^X$. These are components
conjugated to $\chi$ and they are contained in the Gauss constraint $\CG_X$.}
They restrict the commutator of two connections (either $\SA$ or $\tSA$, it does not matter)
\be
I_{(Q)Z}^X \{ \SA_i^Z,\SA_j^W\}_D I_{(Q)W}^Y=0.
\plabel{IQAIQA}
\ee
Since this relation holds for any $\im$, taking into account the result \Ref{commAA},
it gives the following set of constraints\footnote{All these
constraints were checked by direct calculations using
the explicit form of $\CK_{ij}^{ZW}$.} on $\CK_{ij}^{ZW}$
\beq
& I_{(Q)Z}^X \CK_{ij}^{ZW} I_{(Q)W}^Y = 0, &
\plabel{KIQIQ} \nonumber \\
& I_{(P)Z}^X \CK_{ij}^{ZW} I_{(P)W}^Y = 0, &
\plabel{KIPIP} \\
& I_{(Q)Z}^X \CK_{ij}^{ZW} I_{(P)W}^T \Pi_T^Y +
\Pi^X_S I_{(P)Z}^S \CK_{ij}^{ZW} I_{(Q)W}^Y = 0. &
\plabel{KIPIQ} \nonumber
\eeq
In particular, they imply that a relation similar to \Ref{IQAIQA} is also valid
\be
I_{(P)Z}^X \{ \SA_i^Z,\SA_j^W\}_D I_{(P)W}^Y=0.
\plabel{IPAIPA}
\ee

Note that the three equations \Ref{KIPIP} are equivalent to the statement that
\be
\CK_{ij}^{XY} = \tilde \CK_{ij}^{XY} + \Pi^X_Z \tilde \CK_{ij}^{ZW} \Pi_W^Y,
\plabel{KQP}
\ee
where $\tilde\CK_{ij}^{XY}=I_{(Q)Z}^X \CK_{ij}^{ZW} I_{(P)W}^Y$.
Substituting this result into \Ref{commAA}, one finds that the commutator 
can be rewritten as follows
\be
\{ \SA^X_i(x), \SA^Y_j(y)\}_D 
= \hf \( \tilde \CK_{ij}^{XY} + \Pi^X_Z \tilde \CK_{ij}^{ZW} \Pi_W^Y \) \delta(x,y).
\plabel{commutAAA}
\ee
Thus, the dependence on the Immirzi parameter completely disappears from the final result.

Taking into account the expression for $\CK_{ij}^{XY}$ from \cite{SAhil},
one can obtain an explicit result for the commutator of two connections based on 
the representation \Ref{commutAAA}.
Denoting the operator acting on the $\delta$-function in the r.h.s. 
by $\CM_{ij}^{XY}$ as in \Ref{commAAA},
one can derive that it is given by
\be
\CM_{ij}^{XY}=\hf \( \Pi^X_{X'} \tilde\CM_{ij}^{X'Y}-\tilde \CM_{ij}^{XY'}\Pi_{Y'}^Y\),
\plabel{CMMM}
\ee
where 
\be
\tilde\CM_{ij}^{XY}(x,y) =  
-\hf\( \CV_{ij}^{XY,l}(x)\p_l^{(x)}+\CV_{ji}^{YX,l}(y)\p_l^{(y)}\)+\CW_{ij}^{XY}
\plabel{tldM} 
\ee
and
\beq
\CV_{ij}^{XY,l} & =& 
f^{XP}_Q\left[\tQ^l_P\left(
(\Qt\Qt)_{ij}I_{(Q)}^{YQ}+\Qt_i^Y\Qt_j^Q-\Qt_j^Y\Qt_i^Q\right)
+\delta_i^l I_{(Q)}^{YQ}\Qt_j^P\right],
\plabel{VVV} \\
\CW_{ij}^{XY} & = &\hf\( \CL_{ij}^{XY}+\CL_{ji}^{YX}\)
+\frac{g_{SS'}}{2}\(I_{(P)}^{XT}\CV_{ij}^{SY,l} +
I_{(P)}^{YT}\CV_{ji}^{SX,l}\)\Qt_n^{S'}  \p_l \tQ^n_T ,
\label{WWW} \\
\CL_{ij}^{XY} &=& f^{PQ}_Z \left[ \Qt_j^{X}\Qt_n^Y\Qt_i^Z+
(\Qt\Qt)_{in}\Qt_j^{X}I_{(Q)}^{YZ}+\Qt_i^Y\Qt_n^{X}\Qt_j^Z
\right. \label{LLL} \\
&& \quad \qquad  - \left.
\Qt_i^Y\Qt_j^{X}\Qt_n^Z+(\Qt\Qt)_{ij}\Qt_n^{X}I_{(Q)}^{YZ}-
\Qt_j^Y\Qt_n^{X}\Qt_i^Z\right] \tQ^l_P\p_l\tQ^n_Q
\nonumber \\
&+& f^{Q}_{ZP} \left[ \Qt_n^Y\Qt_j^P+
(\Qt\Qt)_{jn}I_{(Q)}^{YP}-\Qt_j^Y\Qt_n^{P}\right]
I_{(Q)}^{ZX}\p_i\tQ^n_Q
+f^Z_{PQ}\Qt_j^{X}\Qt_i^QI_{(Q)}^{YP}\p_l\tQ_{Z}^l .
\nonumber
\eeq
Since this operator is implied to act on $\delta(x,y)$,
the argument of the last term in \Ref{tldM} is not important. 
The antisymmetry of the bracket is ensured by the antisymmetry property of the matrix 
\Ref{VVV}
\be
\CV_{ij}^{XY,l} = - \CV_{ji}^{YX,l} ,
\label{symV}
\ee
which can be checked by straightforward calculations.

\section{The \SLC\ connection of the 3d hypersurface}
\label{C}

In this appendix we demonstrate that the quantity (\ref{gam}), 
which is the non-dynamical part of the
Lorentz connection $\tSA$, has a close relation to the Levi-Civita connection of the
spacelike hypersurface.  
To establish this relation, let us define a connection
$\Gamma^Y_{iX}$ by the condition that all fields $\tQ_X^i$,
$\tP^i_X$ and their inverse are covariantly constant. For example,
\be
\nabla_i\tQ^k_X=\p_i\tQ^k_X+\Gamma^{(0)k}_{ij}\tQ^j_X-
\Gamma^Y_{iX}\tQ^k_Y+\frac{1}{2}\tQ^k_X\log g=0, \plabel{cd}
\ee
where $\Gamma^{(0)k}_{ij}$ is the Levi-Civita connection of 
the metric induced on the 3-dimensional hypersurface, which was defined in
\Ref{met3}, and $g$ is the determinant of this metric \Ref{detg}. 
The last term in eqn. (\ref{cd}) comes from the weight of the field $\tQ^k_X$. 
From these conditions one can show that $\Gamma^X_{iY}$ takes the form
\be
\Gamma^X_{iY}=\tilde\Gamma^X_{iY}-\Pi^X_Z\tilde\Gamma^Z_{iW}\Pi^W_Y,
\ee
where  
\beq
\tilde \Gamma^{X}_{iY}&=& \Qt_k^X I_{(Q)Y}^{W} \p_i \tQ^k_W
+\Qt^X_k\tQ^{j}_Y\Gamma^{(0)k}_{ij}-\frac{1}{2}I_{(Q)Y}^{X}\p_i \log g 
\nonumber \\
&=&\frac{1}{2} g_{YY'}(V^{WXY',l}_{ij}-V^{WY'X,l}_{ij})\p_l\tQ^j_W
\eeq
and $V_{ij}^{YPQ,l}$ denotes the expression in brackets in \Ref{VVV} so that 
$\CV_{ij}^{XY,l}=f^{X}_{PQ}V_{ij}^{YPQ,l}$
The first equality is the direct consequence of eqn. (\ref{cd}) while the second 
is the result of expressing the metric and $\Gamma^{(0)k}_{ij}$ in terms of 
the fields $\tQ$ via \Ref{met3}.

On the other hand, it is easy to check that \Ref{gam} can be rewritten as
\be
\Gamma_i^X=-\frac{1}{2} \CV_{ji}^{WX,l} \,\p_l \tQ_W^j.
\ee
Then using the property \Ref{symV}, one finds
the following relation between $\Gamma_i^X$ and $\Gamma^X_{iY}$
\be
\Gamma_i^X=\frac{1}{2}\CV_{ij}^{XW,l}\p_l \tQ_W^j =
-\frac{1}{2}f^{XY}_{Z}\tilde\Gamma^{Z}_{iY}=
-\frac{1}{4}f^{XY}_{Z}\Gamma^{Z}_{iY}.
\plabel{cc}
\ee
This allows to identify the non-dynamical part of the shifted connection $\tSA_i^X$
with the induced \SLC\ connection on the 3-dimensional slice.


\begin{thebibliography}{99}


\bibitem{Carlip}
S. Carlip, ``Quantum Gravity: a Progress Report",
Rept. Prog. Phys. {\bf 64}, 885 (2001)
[gr-qc/0108040].


\bibitem{loops1}
C.~Rovelli and L.~Smolin,
``Knot theory and quantum gravity,"
Phys.\ Rev.\ Lett.\ {\bf 61}, 1155 (1988).

\bibitem{loops2}
C.~Rovelli and L.~Smolin,
``Loop representation of quantum general relativity,"
 Nucl.\ Phys.\ B {\bf 331}, 80 (1990).

\bibitem{Oriti}
D.~Oriti, ``Spacetime geometry from algebra:
spin foam models for non-perturbative quantum gravity,"
Rept.\ Prog.\ Phys.\ {\bf 64}, 1489 (2001)
[gr-qc/0106091].

\bibitem{perez}
A.~Perez, ``Spin foam models for quantum gravity,"
[gr-qc/0301113].

\bibitem{Rov}
C.~Rovelli, ``Loop Quantum Gravity,"
Living Rev. Rel. {\bf 1}, 1 (1998)
[gr-qc/9710008].

\bibitem{Rov-dif}
M.~Gaul and C.~Rovelli,
``Loop Quantum Gravity and the Meaning of Diffeomorphism Invariance,"
Lectures given at the 35th Karpacz Winter School on Theoretical Physics:
From Cosmology to Quantum Gravity (1999)
[gr-qc/9910079].

\bibitem{BC}
J.W.~Barrett and L.~Crane,
``A Loerntzian signature model for quantum general relativity,"
Class. Quantum Grav. {\bf 17}, 3101 (2000)
[gr-qc/9904025].

\bibitem{sfLor1}
A.~Perez and C.~Rovelli,
``Spin foam model for Lorentzian General Relativity,"
Phys.\ Rev.\ D {\bf 63}, 041501 (2001)
[gr-qc/0009021].

\bibitem{sfLor2}
A.~Perez and C.~Rovelli,
``$3+1$ spinfoam model of quantum gravity with spacelike and timelike
components,"
Phys.\ Rev.\ D {\bf 64}, 064002 (2001)
[gr-qc/0011037].

\bibitem{Maran3}
S.K.~Maran,
``Relating spin foams and canonical quantum gravity:
$(n-1)+1$ formulation of $nD$ spin foams,"
[gr-qc/0412011].
\bibitem{Maran1}
S.K.~Maran,
``A discrete evolution formulation of the spin foam models of BF theory
and gravity,"
[gr-qc/0311090].
\bibitem{Maran2}
S.K.~Maran,
``Ashtekar formulation with temporal foliations,"
[gr-qc/0312111].

\bibitem{imir}
G.~Immirzi,
``Quantum gravity and Regge calculus,"
Nucl.\ Phys.\ Proc.\ Suppl.\ {\bf 57}, 65 (1997)
[gr-qc/9701052].

\bibitem{area1}
C.~Rovelli and L.~Smolin,
{\it Discreteness of area and volume in quantum gravity.} ---
Nucl. Phys. B {\bf 442}, 593 (1995).
\bibitem{ALarea}
A.~Ashtekar and J.~Lewandowski,
``Quantum theory of gravity I: Area operators,"
Class.\ Quantum\ Grav. {\bf 14}, A55 (1997)
[gr-qc/9602046].

\bibitem{SA}
S.~Alexandrov,
``SO(4,C)-covariant Ashtekar--Barbero gravity and
the Immirzi parameter,"
Class.\ Quantum\ Grav.\ {\bf 17}, 4255 (2000)
[gr-qc/0005085].

\bibitem{AV}
S.~Alexandrov and D.~Vassilevich,
``Area spectrum in Lorentz covariant loop gravity,"
Phys.\ Rev.\ D {\bf 64}, 044023 (2001) [gr-qc/0103105].

\bibitem{SAcon}
S.~Alexandrov,
``On choice of connection in loop quantum gravity,"
Phys.\ Rev.\ D {\bf 65}, 024011 (2002)
[gr-qc/0107071].

\bibitem{SAhil}
S.~Alexandrov,
``Hilbert space structure of covariant loop quantum gravity,"
Phys. Rev. D {\bf 66}, 024028 (2002)
[gr-qc/0201087].

\bibitem{AlLiv}
S.~Alexandrov and E.R. Livine,
``SU(2) loop quantum gravity seen from covariant theory),"
Phys. Rev. D {\bf 67}, 044009 (2003)
[gr-qc/0209105].

\bibitem{psn}
E.R. Livine,
``Projected Spin Networks for Lorentz connection:
Linking Spin Foams and Loop Gravity,"
Class. Quantum Grav. {\bf 19}, 5525 (2002)
[gr-qc/0207084].

\bibitem{FrLiv}
L. Freidel and E.R. Livine,
``Spin Networks for Non-Compact Groups,"
J. Math. Phys. {\bf 44}, 1322 (2003)
[hep-th/0205268].

\bibitem{noncomp}
A.~Okolow,
``Hilbert space built over connections with a non-compact structure group,"
[gr-qc/0406028].

\bibitem{BFhigher}
L. Freidel, K. Krasnov, and R. Puzio,
``BF description of higher dimensional gravity theories,"
Adv. Theor. Math. Phys. {\bf 3}, 1289 (1999)
[hep-th/9901069].
 
\bibitem{simplespin}
L. Freidel and K. Krasnov,
``Simple Spin Networks as Feynman Graphs,"
J. Math. Phys., {\bf 41}, 1681 (2000) 
[hep-th/9903192].

\bibitem{barack}
A.O. Baruth and R. Raczka
``Theory of Group Representations and Applications"
World Scientific, 1986.

\bibitem{ruhl}
W. R\"uhl, ``The Lorentz Group and Harmonic Analysis"
W.A Benjamin Inc., New York, 1970.


\bibitem{FLR}
L.~Freidel, R.~E.~Livine and C.~Rovelli,
``Spectra of Length and Area in $(2+1)$ Lorentzian Loop Quantum Gravity,"
[gr-qc/0212077].

\bibitem{thooft1}
G. 't Hooft, ``The evolution of gravitating point particles in $(2+1)$-dimensions,"
Class. Quantum Grav. {\bf 10}, 1023 (1993).

\bibitem{thooft2}
G. 't Hooft, ``Canonical quantization of gravitating point particles 
in $2+1$ dimensions,"
Class. Quantum Grav. {\bf 10}, 1653 (1993)
[gr-qc/9305008].

\bibitem{thooftst}
H.J.~Matschull and M.~Welling,
``Quantum mechanics of a point particle in 2+1 dimensional gravity,''
Class. Quant. Grav. {\bf 15}, 2981 (1998)
[gr-qc/9708054].

\bibitem{zoltan}
Z. Kadar, 
``Polygon model from first order gravity,"
[gr-qc/0410012].

\bibitem{frlo} 
L. Freidel and D. Louapre,
``Ponzano-Regge model revisited. I: Gauge fixing, observables and interacting spinning
particles,"
Class.\ Quant.\ Grav.\  {\bf 21}, 5685 (2004)
[hep-th/0401076].

\bibitem{SAbh}
S. Alexandrov, 
``On the counting of black hole states in loop quantum gravity,"
[gr-qc/0408033]

\bibitem{canBF}
E. Buffenoir, M. Henneaux, K. Noui, and Ph. Roche, 
``Hamiltonian analysis of Plebanski theory,"
Class. Quant. Grav. {\bf 21}, 5203 (2004) 
[gr-qc/0404041].


\end{thebibliography}
\end{document}